\documentclass[aps,prl,twocolumn,superscriptaddress]{revtex4}

\usepackage{graphicx}
\usepackage{bm}
\usepackage{amsmath}

\begin{document}

\title{Terahertz metamaterial with asymmetric transmission}

\author{R. Singh} \email[Email: ]{ranjan.ranjansingh@gmail.com} \affiliation{School of Electrical and Computer Engineering, Oklahoma State University, Stillwater, Oklahoma 74078, USA}

\author{E. Plum} \affiliation{Optoelectronics Research Centre, University of Southampton, SO17 1BJ, UK}

\author{C. Menzel} \affiliation{Institute of Condensed Matter Theory and Solid State Optics, Friedrich Schiller University Jena, Jena 07743, Germany}

\author{C. Rockstuhl} \affiliation{Institute of Condensed Matter Theory and Solid State Optics, Friedrich Schiller University Jena, Jena 07743, Germany}

\author{A. K. Azad} \affiliation{Los Alamos National Laboratory, MPA-CINT, MS K771, Los Alamos, New Mexico 87545, USA}

\author{R. A. Cheville} \affiliation{School of Electrical and Computer Engineering, Oklahoma State University, Stillwater, Oklahoma 74078, USA}

\author{F. Lederer} \affiliation{Institute of Condensed Matter Theory and Solid State Optics, Friedrich Schiller University Jena, Jena 07743, Germany}

\author{W. Zhang} \affiliation{School of Electrical and Computer Engineering, Oklahoma State University, Stillwater, Oklahoma 74078, USA}

\author{N. I. Zheludev}
\affiliation{Optoelectronics Research Centre, University of
Southampton, SO17 1BJ, UK}

\date{\today}

\begin{abstract}
We show for the first time that a planar metamaterial, an array of
coupled metal split-ring resonators with a unit cell lacking mirror
symmetry, exhibits asymmetric transmission of terahertz radiation
(0.25-2.5~THz) propagating through it in opposite directions. This
intriguing effect, that is compatible with Lorentz reciprocity and
time-reversal, depends on a directional difference in conversion
efficiency of the incident circularly polarized wave into one of
opposite handedness, that is only possible in lossy low-symmetry
planar chiral metamaterials. We show that asymmetric transmission is
linked to excitation of enantiomerically sensitive plasmons, these
are induced charge-field excitations that depend on the mutual
handedness of incident wave and metamaterial pattern. Various bands
of positive, negative and zero phase and group velocities have been
identified indicating the opportunity to develop polarization
sensitive negative index and slow light media based on such
metamaterials.
\end{abstract}


\maketitle

In contrast to three-dimensionally chiral structures (e.g. helices),
planar chiral patterns (e.g. flat spirals) have the intriguing
property that their sense of twist is reversed for observation from
opposite sides. Not only human observers, but also circularly
polarized waves incident on opposite sides of a planar chiral
structure, see materials of opposite handedness. It has recently
been discovered that planar chiral metamaterial patterns can show
different levels of total transmission for circularly polarized
waves of the same handedness propagating in opposite directions. The
effect, which has been detected in microwave
\cite{PRL_Fedotov_2006_AsymmetricTransmissionMW,
APL_Plum_2009_2DchiralASR, JOPA_Plum_2009_ExtrinsicChirality,
OL_Zhukovsky_2009_EllipDichroism} and photonic
\cite{NanoLett_Fedotov_2007_AsymmetricTransmission,
NanoLett_Schwanecke_2008_AsymTrans} metamaterials and plasmonic
nanostructures \cite{OE_Drezet_2008_asymNanostruct}, is known as
\emph{asymmetric transmission}. Such asymmetric transmission
phenomenon has not yet been observed for terahertz radiation. The
terahertz spectral region has tremendous technological importance
since many biological materials and substances have molecular
vibration frequencies in this regime, making it highly attractive
for sensing, material characterization, spectroscopy and biomedical
imaging. In spite of intense research activity in this domain over
the past decade terahertz radiation has proved to be extremely
challenging to detect, measure, propagate and manipulate since
electronic and magnetic responses of natural materials die out at
these frequencies, thus earning the name of the so-called
``terahertz gap". Recently, terahertz metamaterials
\cite{Science_Yen_2004_THzMagnetic, Science_Linden_2004_THzMagnetic,
PRL_Moser_2005_THzMetamaterial, OL_Azad_2006_THzSRR,
Nature_2006_THzActiveMM, OE_Paul_2008_THzBulkNIM,
PRL_Zhang_2009_THzChiralNIM, OE_Singh_2008_THzMM,
OL_Singh_2008_THzMM, OE_OHara_2008_THzMMsensing,
AdvMater_Liu_2008_coupledSR} have shown potential for use in the
terahertz gap with their fascinating novel properties but the region
still suffers from a severe shortage of devices needed for fully
exploiting the attractive potential applications of terahertz
radiation.

\begin{figure}[t!]
\includegraphics[width=85mm]{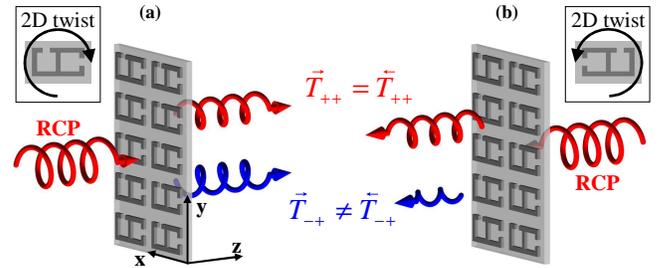}
\caption{\label{fig-concept}(color online). Asymmetric total
transmission of a circularly polarized wave incident on (a) front
and (b) back side of a planar chiral metamaterial. The right-handed
circularly polarized wave (red spiral) is partially converted to the
left-handed (blue spiral) polarization when propagating through the
metamaterial. The conversion efficiency differs for opposite
directions of wave propagation resulting in different levels of
total transmission. Insets show the 2D-chiral twist of the unit
cell, as perceived by the incident wave.}
\end{figure}

In this Letter we report the first experimental observation of
asymmetric transmission in the terahertz domain. We demonstrate a
new type of polarization sensitive terahertz metamaterial device
showing directionally asymmetric transmission of circularly
polarized waves between 0.25 and 2.5~THz. The phenomenon resembles
the non-reciprocal Faraday effect in magnetized media, but takes
place in absence of any magnetic field. Experimentally and
numerically we show that the total transmission level of circularly
polarized waves through a planar chiral metamaterial pattern depends
on both the wave's handedness and propagation direction. Unlike the
Faraday effect in which the asymmetry applies to the transmission
and retardation of the incident circularly polarized wave itself,
asymmetric transmission is a completely reciprocal phenomenon
arising from partial conversion of the incident circularly polarized
wave into one of the opposite handedness. As illustrated by Figure
\ref{fig-concept}, it is the efficiency of circular polarization
conversion that depends on the incident wave's handedness and
propagation direction, and which gives rise to the asymmetry in
total transmission.

\begin{figure}[t!]
\includegraphics[width=85mm]{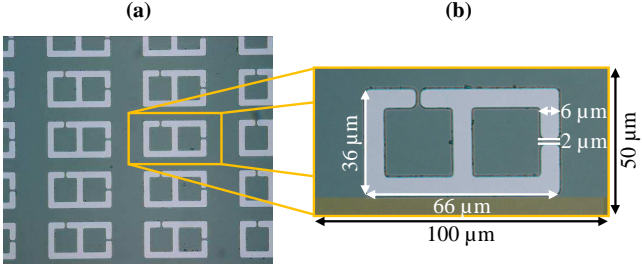}
\caption{\label{fig-structure}(color online). (a) Front side of the
planar chiral terahertz metamaterial consisting of 200~nm thick
aluminum wires on an n-type silicon substrate. (b) Metamaterial unit
cell.}
\end{figure}

The metamaterial structure, which is shown in Figure
\ref{fig-structure}(a), is based on pairs of split rings of
orthogonal orientation that are joined together forming a 2D-chiral
pattern. The planar twist of the structure can be defined as from
``gap on long side" to ``gap on short side", making the structure
right-handed when observed from the structured front and left-handed
when observed from the back, see insets of Figure \ref{fig-concept}.
The planar chiral metamaterial sample was fabricated by conventional
photolithography from a 200~nm thick aluminum layer deposited on a
$640~\mu\text{m}$ thick silicon substrate with n-type resistivity
12~$\Omega$~cm and an absorption constant of 5/cm
\cite{PRB_Singh_2009_THzMM}. Figure \ref{fig-structure}(b) shows
detailed dimensions of the metamaterial's rectangular unit cell
which is $100 \times 50 \mu \text{m}^2$ in size rendering the
structure non-diffracting at normal incidence for frequencies up to
3 THz. We studied the structure using terahertz time-domain
spectroscopy (THz-TDS)
\cite{JOSAB_Grischkowsky_1990_THzSpectroscopy,
OptCommun_He_2006_THzSpectroscopy}. The terahertz beam incident on
the sample had a frequency independent diameter of 3.5~mm and thus
illuminated about 2000 unit cells at the center of the $10 \times
10~\text{mm}^2$ metamaterial array. Using parallel or crossed linear
polarizers placed before and after the sample, we measured all
components of the metamaterial's transmission matrix
$E_i^{\text{trans}} = \tau_{ij} E_j^{\text{\text{inc}}}$, which
relates the incident and transmitted electric fields in terms of
linearly polarized components \cite{PRL_Zhang_2009_THzChiralNIM}.
Amplitude
$|\tau_{ij}(\omega)|=|E_{ij}^{\text{sample}}(\omega)|/|E^{\text{ref}}(\omega)|$
and phase $\arg(\tau_{ij}(\omega))=\angle
[E_{ij}^{\text{sample}}(\omega)/E^{\text{ref}}(\omega)]$ of the
transmission matrix elements were calculated from transmission
measurements taken on the metamaterial
$E_{ij}^{\text{sample}}(\omega)$ with a blank silicon substrate
$E^{\text{ref}}(\omega)$ used as a reference. In order to study the
effect of asymmetric transmission, which occurs for circularly
polarized waves, we transformed the transmission matrix $\tau_{ij}$
from the linear polarization basis to the circular polarization
basis

\begin{eqnarray}
t &=& \begin{pmatrix}t_{++} & t_{+-} \\ t_{-+} & t_{--}
\end{pmatrix} \nonumber \\
&=& \frac{1}{2}
\begin{pmatrix}\tau_{xx}+\tau_{yy}+i(\tau_{xy}-\tau_{yx})&\tau_{xx}-\tau_{yy}-i(\tau_{xy}+\tau_{yx})\\
\tau_{xx}-\tau_{yy}+i(\tau_{xy}+\tau_{yx})&\tau_{xx}+\tau_{yy}-i(\tau_{xy}-\tau_{yx})\end{pmatrix}.
\nonumber
\end{eqnarray}

Transformed in this way the transmission matrix
$E_i^{\text{trans}}=t_{ij}E_j^{\text{inc}}$ directly relates the
incident and transmitted electric fields in terms of right-handed
(RCP, +) and left-handed (LCP, -) circularly polarized components,
while the squares of its elements $T_{ij}=|t_{ij}|^2$ correspond to
transmission and circular polarization conversion in terms of power.
The metamaterial's transmission characteristics, as well as the
current configurations excited in the metamaterial unit cell, were
also simulated using the Fourier Modal Method (FMM)
\cite{JOSAA_Li_1997_FMM}.

\begin{figure}[t!]
\includegraphics[width=85mm]{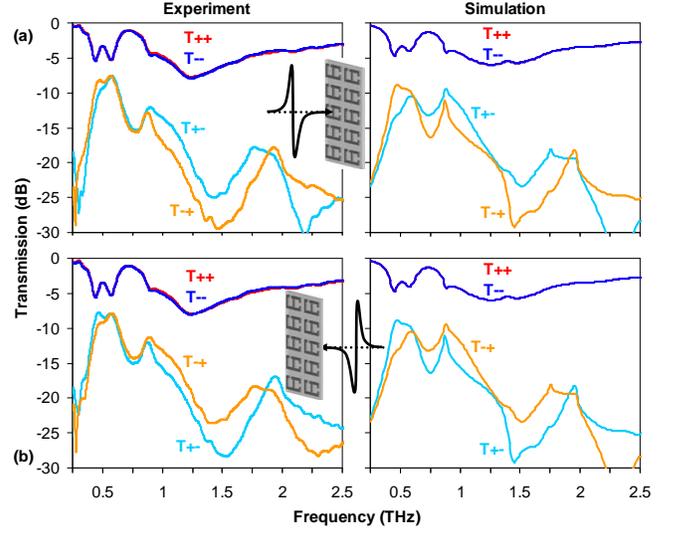}
\caption{\label{fig-transmission}(color online). Transmission
spectra for circularly polarized terahertz waves incident on (a)
front and (b) back of the metamaterial array. It can be clearly seen
that the circular polarization conversion efficiencies $T_{-+}$ and
$T_{+-}$ are reversed for opposite directions of propagation.}
\end{figure}

As illustrated by Figure \ref{fig-transmission}, our numerical and
experimental results show that the metamaterial's direct
transmission for circular polarization is reciprocal as coefficients
$t_{++}=t_{--}$ are both identical and independent of the direction
of propagation. Thus optical activity $\arg(t_{++})-\arg(t_{--})$
and circular dichroism $T_{++}-T_{--}$, which are associated with
three-dimensional chirality, are negligible indicating that the
metamaterial - which is formally 3D-chiral due to the substrate on
only one side of the metal pattern
\cite{APL_Menzel_2008_ParameterRetrieval} - behaves like a truly
planar structure. Furthermore the fact that $t_{++}$ and $t_{--}$ do
not depend on the direction of propagation demonstrates complete
absence of the Faraday effect.

In contrast to direct transmission, the right-to-left $T_{-+}$ and
left-to-right $T_{+-}$ circular polarization conversion levels
depend on both the direction of wave propagation and the handedness
of the incident wave, indicating the presence of the asymmetric
transmission effect. Importantly counter-propagating circularly
polarized waves of the same handedness experience different levels
of circular polarization conversion, while their direct transmission
levels are identical, for example $\overrightarrow{T}_{-+}\neq
\overleftarrow{T}_{-+}$ and $\overrightarrow{T}_{++}=
\overleftarrow{T}_{++}$ in case of RCP. It follows that the
metamaterial's total transmission for RCP, defined as $T_+ =
T_{++}+T_{-+}$, is asymmetric with respect to opposite directions of
wave propagation. Furthermore it must be noted that the conversion
efficiencies for RCP and LCP are simply interchanged for opposite
directions of wave propagation, i.e.
$\overrightarrow{T}_{+-}=\overleftarrow{T}_{-+}$. This has two
significant consequences: Firstly, the directional transmission
asymmetry $\overrightarrow{T}_+
-\overleftarrow{T}_+=\overrightarrow{T}_{-+}-\overleftarrow{T}_{-+}$
is identical to the total transmission difference for opposite
circular polarizations propagating in the same direction,
$\overrightarrow{T}_+
-\overrightarrow{T}_-=\overrightarrow{T}_{-+}-\overrightarrow{T}_{+-}$.
Secondly, the metamaterial has the same transmission properties for
circularly polarized waves of opposite handedness propagating in
opposite directions, i.e. $\overrightarrow{T}_+=\overleftarrow{T}_-$
and $\overrightarrow{T}_-=\overleftarrow{T}_+$.

\begin{figure}[t!]
\includegraphics[width=85mm]{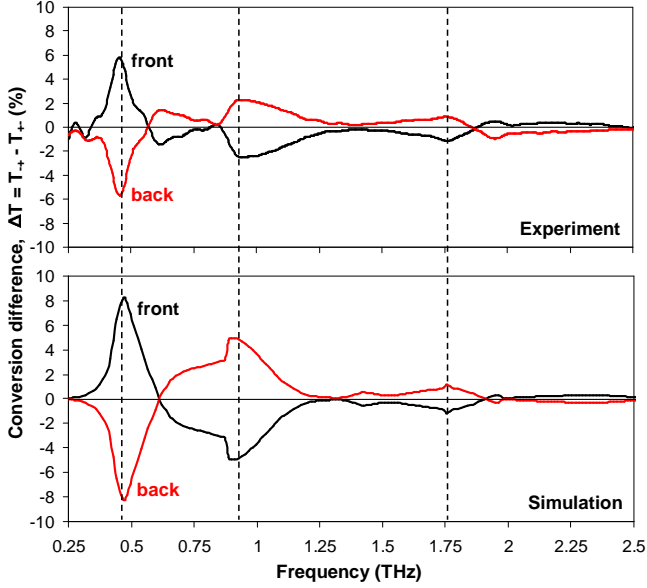}
\caption{\label{fig-asymmetry}(color online). Transmission
difference $\Delta T = T_+ - T_- = T_{-+} - T_{+-}$ for right-handed
and left-handed circularly polarized waves incident on the front
(black curves) or back (red curves) of the metamaterial sample.}
\end{figure}

Figure \ref{fig-asymmetry} shows the total transmission asymmetry
for circularly polarized terahertz waves incident on the structure's
front and back directly. Experimental and numerical results are
generally in good agreement and show that asymmetric transmission
takes place over the entire studied spectral range from 0.25 to
2.5~THz. The largest asymmetry of total transmission occurs around
0.47~THz, where the structure is measured to be 6\% (simulation:
8\%) more transparent for RCP than LCP terahertz waves incident on
its front. For waves incident on the metamaterial's back the
situation is reversed with larger total transmission for LCP than
RCP by the same amount.

\begin{figure}[t!]
\includegraphics[width=85mm]{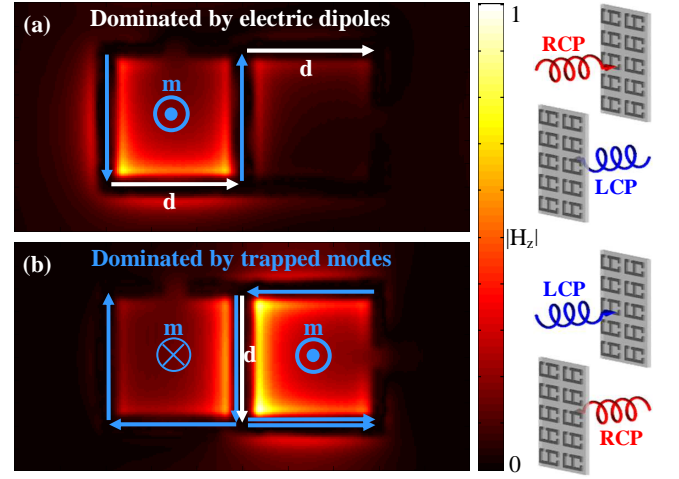}
\caption{\label{fig-simulation}(color online). Enantiomerically
sensitive plasmons linked to the resonant transmission asymmetry at
0.47~THz. The current oscillations in the wires of the structure are
represented by arrows, while the color-scale indicates the magnitude
of the magnetic field the currents induce normal to the
metamaterial's plane. Note the radical difference in the excitation
patterns excited by circular polarizations of either opposite
handedness or opposite propagation direction.}
\end{figure}

The resonant character of asymmetric transmission at 0.47~THz is
linked to the excitation of enantiomerically sensitive plasmons,
these are induced charge-field excitations that depend on the mutual
handedness of the wave and the metamaterial pattern
\cite{NanoLett_Fedotov_2007_AsymmetricTransmission,
APL_Plum_2009_2DchiralASR}. Indeed, numerical simulations show
radically different patterns of currents when the metamaterial
structure is excited by left or right circularly polarized waves: a
RCP wave entering the metamaterial from the front side induces a
strongly anisotropic electric dipole current oscillation $d$ along
the long side of the unit cell that is responsible for the efficient
circular polarization conversion, see Figure
\ref{fig-simulation}(a). On the contrary, the current mode excited
by an LCP wave propagating in the same direction is dominated by
high Q-factor anti-symmetric current oscillations, which correspond
to magnetic moments $m$ oscillating normal to the metamaterial
plane, see Figure \ref{fig-simulation}(b). As the magnetic
components cannot interact with the incident and scattered fields
(which propagate parallel to $m$), this current configuration is
weakly coupled to free space. These high Q-factor currents, known as
trapped or closed modes \cite{PRL_Fedotov_2007_TrappedModes,
APL_Plum_2009_2DchiralASR} are responsible for the smaller level of
circular polarization conversion resulting in lower resonant total
transmission.

\begin{figure}[t!]
\includegraphics[width=85mm]{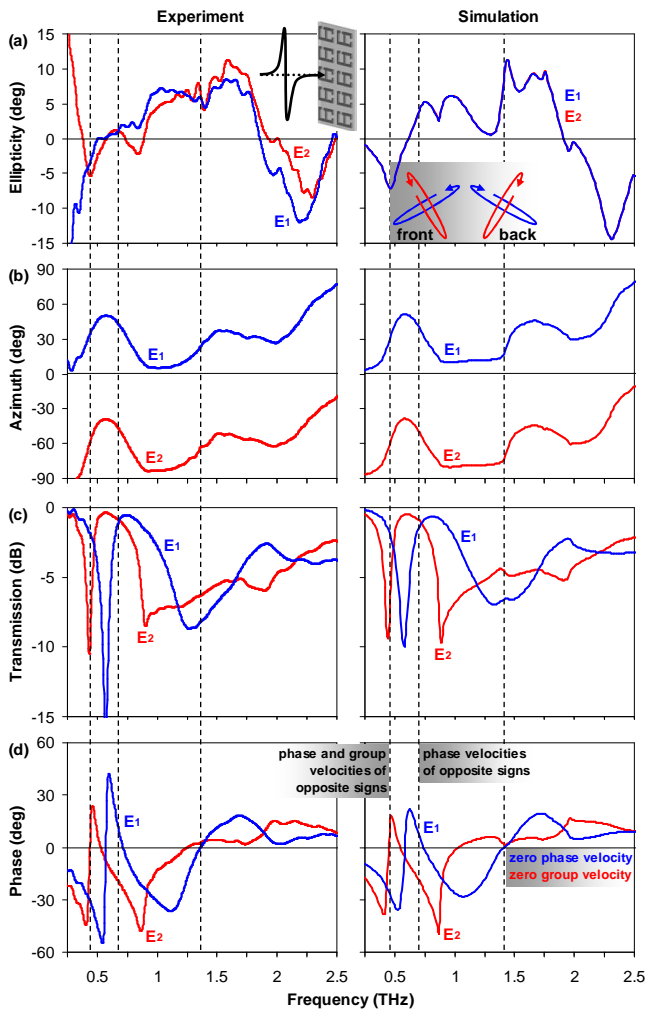}
\caption{\label{fig-eigenstates}(color online). Transmission
eigenstates $E_1$ and $E_2$ for forward propagation in terms of (a)
ellipticity angle and (b) azimuth. (c) Transmission level and (d)
phase delay for these eigenpolarizations.}
\end{figure}

Although the metamaterial shows strong circular polarization
conversion, certain polarization states remain unchanged on
transmission. Ellipticity and azimuth of the transmission
eigenstates for waves incident on the structure's front are shown in
Figures \ref{fig-eigenstates} (a) and (b) respectively. In contrast
to optical activity, or the Faraday effect, asymmetric transmission
is associated with co-rotating elliptical eigenstates. The
eigenpolarizations have orthogonal orientations and their handedness
is reversed for the opposite propagation direction, see inset to
Figure \ref{fig-eigenstates} (a). Figures \ref{fig-eigenstates} (c)
and (d) illustrate the metamaterial's transmission properties for
its eigenpolarizations in terms of transmission levels and phase
delay. Intriguingly, the metamaterial pattern can introduce positive
as well as negative phase delays, indicating that positive and
negative phase velocities should be expected in a bulk material
based on the structure. The group velocity, which is proportional to
the slope of the phase dispersion, can only be discussed for stable
eigenstates. However, eigenstate stability in a finite medium may be
achieved for any frequency by limiting the pulse spectrum.
Therefore, in principle, also the group velocity may be defined even
if the eigenstates depend on frequency. Various bands of positive
and negative phase dispersion indicate that in 2D-chiral bulk
metamaterials group velocities of either sign may be possible. For
example at 0.45~THz the eigenstates appear to have both opposite
phase velocities and opposite group velocities, while at 0.95~THz
their group velocities are almost identical, but their phase
velocities have opposite signs. Finally at 1.41~GHz the phase
velocity of eigenstate $E_1$ and the group velocity of $E_2$ are
zero. These results indicate an opportunity to develop polarization
sensitive negative index and slow light media for elliptically
polarized waves on the basis of bulk 2D-chiral anisotropic
metamaterials.

In conclusion, we present the first experimental and numerical
evidence of asymmetric transmission of circularly polarized
terahertz waves through a planar chiral metamaterial. The observed
effect is due to different levels of circular polarization
conversion for waves incident on the planar structure's front and
back and may lead to a novel class of polarization sensitive
terahertz devices such as polarization and direction sensitive beam
splitters, circulators and sensor components.

\begin{acknowledgments}
Financial support of the U.S. National Science Foundation, the
Engineering and Physical Sciences Research Council, UK, the European
Community project ENSEMBLE and the German Federal Ministry of
Education and Research project Metamat are acknowledged.
\end{acknowledgments}


\end{document}